\documentclass{emulateapj}

\usepackage{graphicx}
\usepackage{amsmath}
\usepackage{latexsym}

\def\H2{{{\rm H}_2}}
\def\apjs{APJS}
\def\beq{\begin{equation}}
\def\eeq{\end{equation}}
\def\bey{\begin{eqnarray}}
\def\eey{\end{eqnarray}}

\def\sun{\odot}
\def\lsim{\mathrel{\raise.3ex\hbox{$<$\kern-.75em\lower1ex\hbox{$\sim$}}}}
\def\gsim{\mathrel{\raise.3ex\hbox{$>$\kern-.75em\lower1ex\hbox{$\sim$}}}}

\newcommand{\be}{\begin{equation}}
\newcommand{\ee}{\end{equation}}
\newcommand{\tev}{\ensuremath{\mathrm{\,Te\kern -0.1em V}}\xspace}
\newcommand{\gev}{\ensuremath{\mathrm{\,Ge\kern -0.1em V}}\xspace}
\newcommand{\tevt}{\ensuremath{\mathrm{Te\kern -0.1em V}}\xspace}
\newcommand{\kev}{\ensuremath{\mathrm{\,ke\kern -0.1em V}}\xspace}
\newcommand{\mev}{\ensuremath{\mathrm{\,Me\kern -0.1em V}}\xspace}

\newcommand{\jcap}{J. Cosmol. Astropart. P.}

\def\HII{{\rm H}_{\rm II}}
\def\HI{{\rm H}_{\rm I}}
\def\H2{{\rm H}_2}

\journalinfo{}
\submitted{submitted to APJ}

\begin{document}

\title{Circum-Galactic Gas and the Isotropic Gamma Ray Background}

\author{Robert Feldmann$^{1,2}$}
\author{Dan Hooper$^{1,2,3}$}
\author{Nickolay Y.~Gnedin$^{1,2,3}$}
\affiliation{$^{1}$Center for Particle Astrophysics, Fermi National Accelerator Laboratory, Batavia, IL 60510, USA; feldmann@fnal.gov}
\affiliation{$^{2}$Kavli Institute for Cosmological Physics, The University of Chicago, Chicago, IL 60637 USA} 
\affiliation{$^{3}$Department of Astronomy and Astrophysics, University of Chicago, Chicago, IL 60637, USA}


\begin{abstract}

Interactions of cosmic rays with the interstellar gas and radiation fields of the Milky Way provide the majority of the gamma rays observed by the Fermi Gamma Ray Space Telescope. In addition to the gas which is densely concentrated along the Galactic Disk, hydrodynamical simulations and observational evidence favor the presence of a halo of hot ($T\sim10^6$ K) ionized hydrogen ($\HII{}$), extending with non-negligible densities out to the virial radius of the Milky Way. We show that cosmic ray collisions with this circum-galactic gas should be expected to provide a significant flux of gamma rays, on the order of 10\% of the observed isotopic gamma ray background at energies above 1 GeV. In addition, gamma rays originating from the extended $\HII$ halos of other galaxies along a given line-of-sight should contribute to this background at a similar level.
\end{abstract}


\maketitle

\section{Introduction}

In addition to emission clearly associated with a Galactic origin, the Fermi Gamma Ray Space Telescope (as well as its predecessors, SAS-2;~\citealt{sas}; and EGRET;~\citealt{egret}) has identified an approximately isotropic background of gamma rays~\citep{fermibg} which is generally considered to be extragalactic in nature. For many years, unresolved blazars were considered likely to be a dominate contributor~\citep{1996ApJ...464..600S, 2005AIPC..745..578K, 2011PhRvD..84j3007A}, although the lack of observed anisotropy in the gamma ray background constrains the blazar contribution to be less than about 20\% of the observed emission~(\citealt{2012PhRvD..85h3007A, 2012arXiv1202.5309C}, see, however,~\citealt{Harding:2012sa}). In light of this, starburst galaxies are now perhaps the most promising class of sources to generate the majority of this observed emission~\citep{Thompson:2006qd}. Other extragalactic sources, including gamma ray bursts~\citep{2007ApJ...656..306C}, shocks associated with large-scale structure formation~\citep{2000Natur.405..156L, 2003ApJ...585..128K}, ultra-high energy cosmic ray interactions~\citep{Kalashev:2007sn}, and dark matter annihilations or decays~\citep{2002PhRvD..66l3502U, 2010PhRvD..81d3505B} have each been considered as possible contributors to the observed background.

And while the roughly isotropic nature of this background is suggestive of an extragalactic origin, it is difficult to exclude the possibility that a significant fraction of this emission is produced more locally. It was suggested by \cite{Keshet:2003xc}, for example, that inverse Compton scattering of high-energy electrons in a halo extending out to at least several tens of kiloparsecs (kpc) beyond the disk of the Milky Way could potentially account for much of the ``extragalactic'' background. In this article, we consider an alternative Galactic contribution to the isotropic gamma ray background. In particular, we consider cosmic ray proton interactions with a spatially extended halo of hot and ionized gas. 

This scenario is particularly plausible for a number of reasons. Firstly, there is an increasingly compelling body of evidence in favor of the presence of an extended circum-galactic reservoir of hot ($T\sim{}10^6$ K) and ionized gas. Observationally, this includes the detection of diffuse soft X-ray emission \citep{2000ApJ...543..195K, 2002A&A...389...93L, 2006ApJ...645...95H}, the presence of absorption lines in the ultraviolet and in X-ray bands consistent with a diffuse background \citep{2003ApJS..146..165S, 2003Natur.421..719N, 2003ASSL..281..109R, 2007ARA&A..45..221B}, H$\alpha$ emission and morphological peculiarities of the Magellanic stream caused by interactions with the circum-galactic gas \citep{1996AJ....111.1156W, 2003ApJ...586..170P, 2007ApJ...670L.109B, 2012MNRAS.421.2109B}, the distorted structure of high velocity clouds \citep{2000A&A...357..120B, 2007ApJ...656..907P, 2011MNRAS.418.1575P}, and the $\HI$ deficiency of Milky-Way dwarf galaxies as a result of ram-pressure stripping \citep{2000ApJ...541..675B, 2009ApJ...696..385G}. And while the spatial extent and total mass of this circum-galactic halo are rather uncertain, sizes of $\sim$$100$ kpc or more are plausible (e.g., \citealt{2000ApJ...541..675B, 2009ApJ...696..385G}, cf. \citealt{2010ApJ...714..320A}). Simulations and semi-analytic models each predict that the circum-galactic halo should extend out to the virial radius, and contains a large fraction or most of the so-called missing baryons \citep{2002MNRAS.335..799T, 2004MNRAS.355..694M, 2006ApJ...644L...1S, 2009ApJ...697...79R, 2010MNRAS.407.1403C, 2011arXiv1108.2271G}. 

Secondly, as only a small fraction of the cosmic rays produced throughout the history of the Milky Way are presently confined to the Galactic disk (see Sec.~\ref{crs}), one expects there to be a spatially extended halo of cosmic rays. For reasonable estimates for the diffusion of cosmic rays in the outer halo of the Milky Way, we find that the rate of interactions between cosmic rays and the circum-galactic gas are sufficient to produce a significant fraction (on the order of 10\%) of the observed isotropic gamma ray background. 

The remainder of this article is structured as follows. In Sec.~\ref{gassec}, we discuss theoretical predictions which favor the existence of an extended halo of hot, ionized gas around the Milky Way and compare them with observational data. In Sec.~\ref{crs}, we discuss the characteristics of the cosmic rays in this outer halo, and in Sec.~\ref{gammasec} we estimate the gamma ray flux from their interactions with the circum-galactic gas. Finally, in Sec.~\ref{conclusions}, we discuss our results and draw conclusions.

\section{The Distribution of Gas in the Outer Halo of the Milky Way}
\label{gassec}

While gas in the form of neutral atomic and molecular hydrogen is strongly concentrated within the disk of the Milky Way, a halo of ionized hydrogen ($\HII$) may extend out to the virial radius of the galaxy. In this section, we discuss estimates of the distribution of this circum-galactic gas, based on a high resolution cosmological, hydrodynamical simulation run with the ART code (Adaptive Refinement Tree; \citealt{1997ApJS..111...73K, 2002ApJ...571..563K}). As the set-up and details of the physical modeling are similar to that of previous numerical experiments performed by some of the authors \citep{2011ApJ...728...88G, 2011ApJ...732..115F}, here we give only a brief recount and highlight the differences between these various simulations.

ART is an Eulerian hydrodynamics + N-body code based on the adaptive mesh refinement technique that allows one to increase the resolution selectively in a specified region of interest, here within five virial radii ($R_{\rm vir}=400$ kpc) around a randomly selected $M_{\rm vir}=2\times{}10^{12}$ $M_\odot$ dark matter halo in an 8.6 Mpc box. The virial mass and radius refer to a region that is 180 times denser than the mean density\footnote{Alternatively, $M_{200}=1.7\times{}10^{12}$ $M_\odot$ and $R_{200}=243$ kpc are the quantities corresponding to a region that is 200 times denser than the critical density of the universe.} of the universe. The selected dark matter halo is embedded in layers of subsequently lower dark matter resolution to reduce the computational cost, while still being able to correctly capture the impact of large scale tidal fields \citep{1991ApJ...368..325K, 2001ApJS..137....1B}. 

The simulation is started from cosmological initial conditions consistent with WMAP data \citep{2003ApJS..148..175S}: $h=0.7$, $\Omega_\Lambda=0.7$, $\Omega_{\rm m}=0.3$, $\Omega_b=0.043$, (implying a universal baryon fraction of $f_{\rm b}\equiv\Omega_b / \Omega_{\rm m} = 0.143$), and $\sigma_8=0.9$. In contrast to the simulations presented by~\cite{2011ApJ...728...88G}, this simulation is continued self-consistently down to $z=0$.

The simulation includes a photo-chemical network to compute abundances of the various hydrogen (including $\H2$) and helium species  \citep{2009ApJ...697...55G, 2011ApJ...728...88G}. It accounts for metal enrichment from type Ia and type II supernovae, but not for any additional thermal or momentum feedback. Metal-dependent radiative cooling rates are computed in the optically thin limit. Radiative transfer of UV radiation from stellar sources is followed in the OTVET approximation \citep{2001NewA....6..437G}. Star formation is based on the amount of molecular hydrogen present.

The simulation has a peak spatial resolution of 250 pc in comoving coordinates. Dark matter particles in the high resolution region have a mass of $1.4\times{}10^{6}$ $M_\odot$, while the masses of stellar particles depend on the star formation rate, see e.g., \cite{2012arXiv1204.3910F}, but are no smaller than $10^3$ $M_\odot$.

\begin{figure}[t]
\centering
\includegraphics[angle=0.0,width=3.2in]{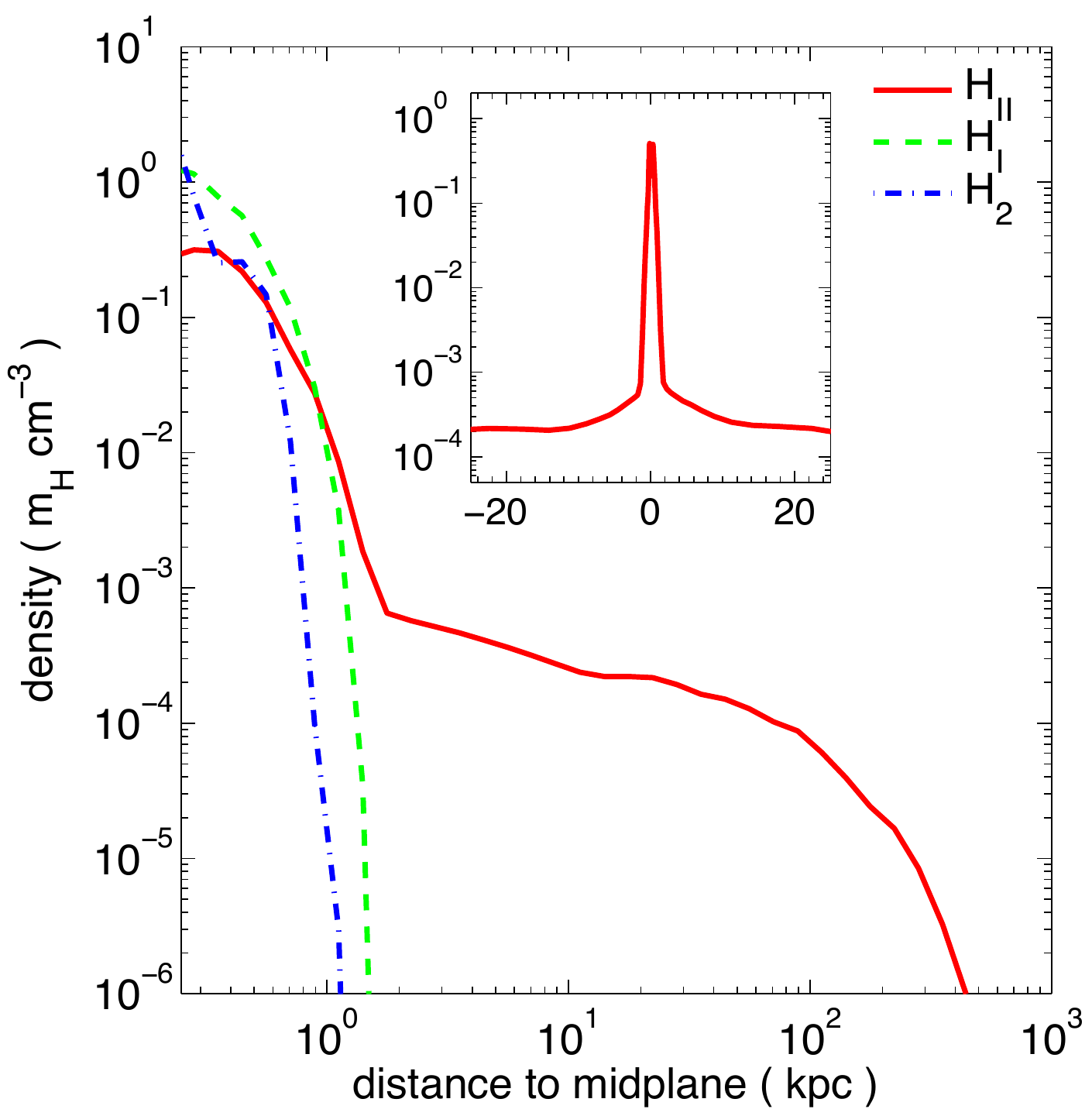}
\caption{The density of ionized hydrogen ($\HII$), neutral hydrogen ($\HI$), and molecular hydrogen ($\H2$) as a function of distance to the Galactic plane, relative to a location 8 kpc from the Galactic center, as found by the simulation described in the text. The inset shows the Reynolds layer of $\HII$ above and below the midplane and part of the extended diffuse circum-galactic component. Note that although the $\HI$ and $\H2$ densities fall off to negligible levels outside of the Galactic plane, a significant density of $\HII$ gas persists to the virial radius of the halo. }
\label{gas}
\end{figure}

At $z=0$, a massive disk galaxy resides at the center of the simulated Milky-Way-like halo. It is immersed in a low density halo of circum-galactic gas that extends out to the virial radius of the halo. In Fig.~\ref{gas}, we show the density profile of ionized, atomic and molecular hydrogen as function of the height above the Galactic plane centered on a midplane annulus with a radius of 8 kpc relative to the Galactic center. Atomic and molecular hydrogen are strongly confined to the Galactic plane and drop to negligible densities $\sim{}1-1.5$ kpc above the midplane. A Reynolds layer of ionized hydrogen surrounds the neutral components of the interstellar medium and extends out to $\sim{}2$ kpc above the midplane. But ionized gas is also present at much larger radii. In fact, the simulation predicts that ionized circum-galactic gas with a density of $\sim 10^{-4}$ cm$^{-4}$ is present out to $\sim 100$ kpc and beyond.

\begin{figure}[t]
\centering
\includegraphics[angle=0.0,width=3.2in]{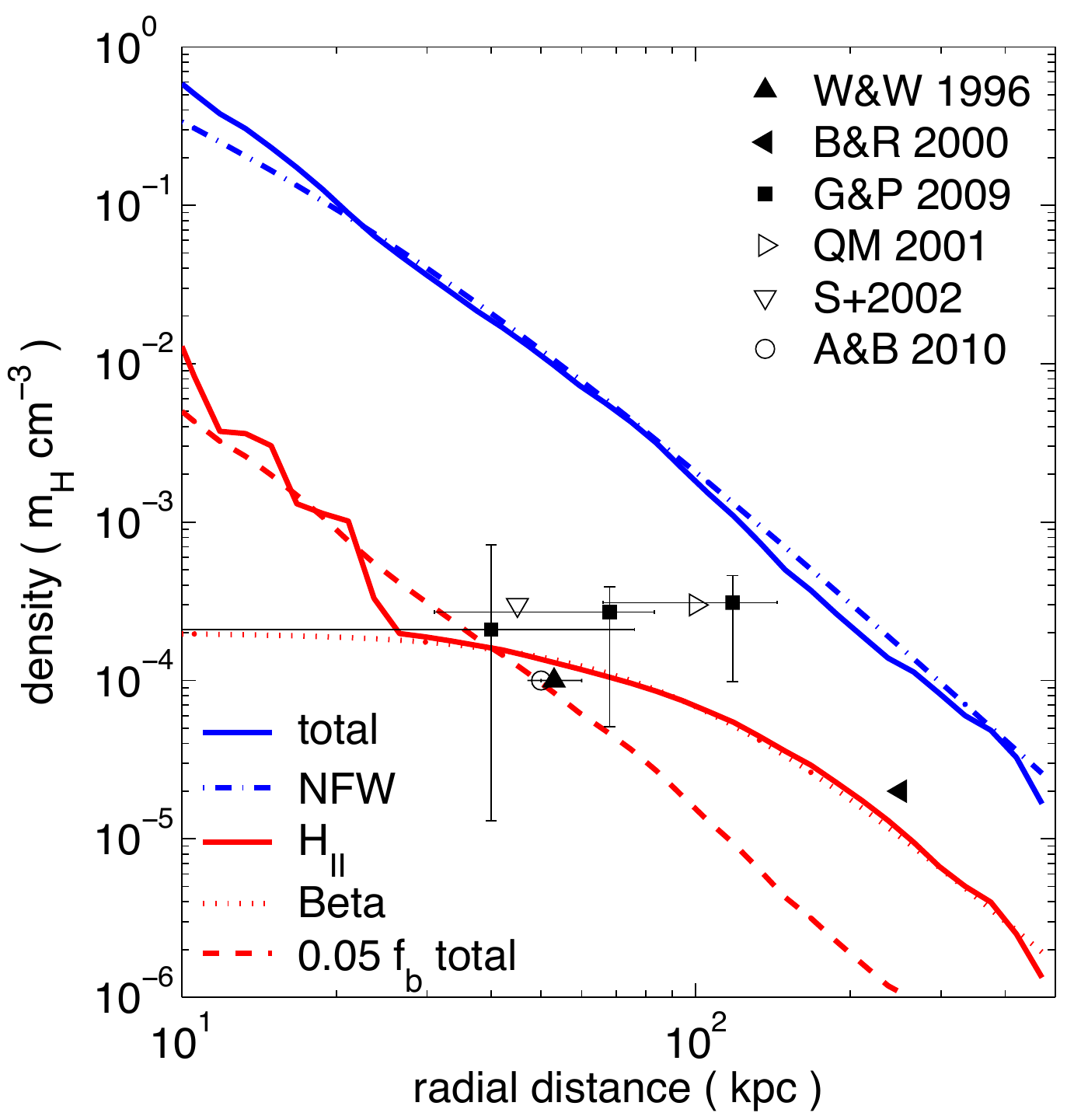}
\caption{Comparison between our simulation and observations. Shown are the total matter density profile as measured in the simulation (blue solid line), the total matter density profile of a dark matter halo with NFW profile and $M_{200}=2\times{}10^{12}$ $M_\odot$ and $R_{200}=250$ kpc (blue dot-dashed line), the profile of $\HII$ as measured in the simulation (red solid line), a fit of a beta-profile to the $\HII$ density profile over the range 30-400 kpc (red dotted line), and the total matter density profile rescaled by a factor 0.05 $f_{\rm b}$ (red dashed line). Note that, in contrast to Fig.~\ref{gas}, density profiles are measured in spherical bins around the center of the galaxy. The data points indicate observational estimates of the circum-galactic gas (filled symbols) or upper limits (empty symbols) \citep{1996AJ....111.1156W, 2000ApJ...541..675B, 2009ApJ...696..385G, 2001ApJ...555L..95Q, 2002ApJ...576..773S, 2010ApJ...714..320A}. The total matter density distribution is well approximated by an NFW profile, except in the central region of the halo ($<20$ kpc), where cooling and adiabatic contraction lead to a density enhancement. The $\HII$ density profile that our simulation predicts is in good agreement with observations. A rescaling of the total matter density profile by a factor 0.05 $f_{\rm b}$ matches the $\HII$ density at $r\sim{}50$ kpc, but severely underestimates the gas density at larger radii and the overall mass in the circum-galactic component. Our simulation predicts that the baryons in the circum-galactic $\HII$ account for $\sim{}0.25$ $f_{\rm b}$ of the total mass in the halo.}
\label{gas2}
\end{figure}

In Fig.~\ref{gas2}, we compare the predictions of this simulation to observations. Note that we switch from cylindrical profiles at the solar circle in Fig.~\ref{gas} to spherical profiles around the galactic center. The predicted density profile of the circum-galactic $\HII$ is consistent with density estimates and upper limits based on observations. We note that the uncertainties in the current observational estimates are substantial (if constrained at all), especially at large radii ($\gtrsim{}100$ kpc). Nonetheless, the agreement is encouraging.

Fig.~\ref{gas2} also shows that the shape of the density profile of the circum-galactic gas is different from the one of the total matter (cf.~\citealt{2010ApJ...714..320A}). While a rescaling of the total matter density profile with a factor of $\sim{}$0.05 $f_{\rm b}$ allows to match the observed density at $\sim{}$50 kpc, it leads to a severe underestimation of the density at larger radii and the overall mass in the circum-galactic  $\HII$ component. In fact, our simulation predicts that the $\HII$ density distribution is reasonably well described by a beta-profile \citep{1985ApJ...293..102F} that falls of more gradually with radius than an NFW profile. After subtracting the $\HII$ mass within the central 25 kpc, the predicted mass of the circum-galactic ionized gas is $2\times{}10^9$ $M_\odot$ within 50 kpc, $1\times{}10^{10}$ $M_\odot$ within 100 kpc, $3.5\times{}10^{10}$ $M_\odot$ within 200 kpc, and $7.3\times{}10^{10}$ $M_\odot$ within $R_{\rm vir}=400$ kpc. The latter mass corresponds to $\sim$$0.25$ $f_{\rm b}$ of the virial mass of the halo and is comparable to the combined mass of the observed stellar and neutral gas components in the Milky Way. Consequently, the baryonic fraction of the Milky Way halo is $\sim{}50\%$ of the universal value $f_{\rm b}$, i.e., half of the baryons must have been expelled from the halo or been prevented from being accreted onto it in the first place\footnote{If the baryon fraction of the Milky Way halo were close to the universal value  $f_{\rm b}$, the circum-galactic gas densities and our predictions for the gamma-ray fluxes would need to be increased by about a factor 3.}.

We note that this result is in line with the detection of a diffuse X-ray halo around the giant spiral galaxy NGC 1961 \citep{2011ApJ...737...22A}. The observationally derived mass of the ionized gas within $\sim$500 kpc is $\sim$$(1-3)\times{}10^{11}$ $M_\odot$. This corresponds\footnote{The inclination and hence total mass of NGC 1961 are rather uncertain. We use the updated inclination estimate $i\sim{}65^\circ$ \citep{2009A&A...503...73C}, which implies a circular velocity of $\sim{}300$ km/s at 34 kpc \citep{2008AJ....135..232H} and thus a halo that is 2-2.5 times more massive than the halo of the Milky Way.} to an $\HII$ mass fraction of $\sim$$0.1-0.35$ $f_{\rm b}$, similar to our prediction of $\sim$$0.25$ $f_{\rm b}$. Interestingly, the picture may be different for elliptical galaxies. Recent observations of X-ray halos around elliptical galaxies of relatively low halo mass (a few $10^{12}$ $M_\odot$) indicate that the fraction of baryons in their ionized gas halos is significantly higher $\sim{}0.5-0.7$ $f_{\rm b}$, which together with their large stellar mass fraction is sufficient to account for essentially all of the ``missing baryons'' \citep{2011ApJ...729...53H, 2012arXiv1204.3095H}.

\section{An Extended Halo of Cosmic Rays}
\label{crs}

In galaxies such as the Milky Way, cosmic rays are thought to originate from sources distributed throughout the disk, such as supernova remnants. These cosmic rays proceed to propagate diffusively through the Galactic magnetic field, resulting in the formation of a steady-state distribution which is present throughout the volume of the disk. Measurements of unstable species in the cosmic ray spectrum can be used to deduce the timescale of how long cosmic rays remain in the disk before escaping. The $^{10}$Be-to-$^9$Be ratio is particularly useful in this regard, as $^{10}$Be is the longest lived and best measured unstable secondary cosmic ray species~\citep{Strong:2007nh}. These measurements lead to an estimate of the cosmic ray escape time of $t_{\rm esc}\approx 2\times 10^7$ years, corresponding to only $\sim$0.2\% of the star formation history of the Milky Way. Therefore only a very small fraction of the cosmic rays produced throughout the history of the Milky Way are presently confined to its disk. 

Under simple 1-D diffusion (such as the case of diffusion away from and perpendicular to the disk), a cosmic ray will traverse a distance $D\sim \sqrt{tK}$ in a time $t$, where $K$ is the diffusion coefficient. Measurements of the local cosmic ray boron-to-carbon ratio imply a value of $K\sim 10^{29}$ cm$^2$/s for $\sim$GeV-TeV protons within the diffusion zone of the Galactic disk. For such a diffusion coefficient, cosmic rays injected since the Milky Way's peak in star formation (a few Gyr ago) will be concentrated within $\sim$30 kpc of the disk. Of course, the magnetic fields are expected to be considerably weaker in the outer halo than within the disk, potentially facilitating much more efficient diffusion. For reasonable estimates of the magnetic fields, however, one still expects a large fraction of the integrated cosmic ray luminosity of the Milky Way to be contained within an extended cosmic ray halo inside of the Galaxy's virial radius. A lower limit on the spatial extent of the Milky Way's cosmic ray halo can be estimated by simply considering the galaxy's power output in cosmic rays. Presently, the total energy in cosmic rays throughout the volume of the disk is $\sim$ 1 eV/cm$^3 \times \pi$(200 pc)(15 kpc)$^2\sim 4 \times 10^{66}$ eV. In comparison, assuming a constant cosmic ray injection rate, the total energy injected into cosmic rays throughout the history of the Milky Way is $\sim 5 \times 10^{40}$ erg/s $\times 10^{10}$ yr $\sim 1.6 \times 10^{58}$ erg $\sim 10^{70}$ eV, which is about 2500 times higher than is currently confined to the disk. Alternatively, one can consider the ratio of the duration of the Milky Way's star formation history to the escape time of cosmic rays in the disk, ($10^{10}$ yr)/($2\times 10^7$ years)$\sim 500$. According to either estimate, the vast majority of cosmic rays generated by our Galaxy are no longer contained it its disk. If we also take into account variations in the star formation rate (which was a factor of a few higher than present rates between roughly 2 and 6 billion years ago), we find that only about $\sim$0.08\% of the cosmic rays created over the history of the Milky Way are currently present within the disk. If we (unrealistically) assume that the cosmic rays are distributed with a uniform density equal to the value in the disk, the population of cosmic rays would fill a sphere approximately $\sim$35 kpc in radius. As measurements of secondary-to-primary ratios tell us that the cosmic ray density must drop off significantly beyond a few kpc from the disk, we conclude that this halo of cosmic rays must be at least $\sim$50 kpc, and possibly much larger, in radial extent.

If the magnetic fields in the outer halo are on the order of $\sim$$10^{2}$ times smaller than those in the disk (or that the diffusion coefficient is $10^{2}$ times larger than in the disk), for example, the majority of the cosmic rays produced throughout the history of the Milky Way will remain within a few hundred kpc of the disk, corresponding to cosmic ray densities in the outer halo on the order of 0.001-0.1 of that found in the disk. Through interactions with the extended halo of ionized gas described in the previous section, these cosmic rays can provide a flux of gamma rays which constitutes a significant fraction of the observed isotropic gamma ray background.

\section{The Contribution to the Isotropic Gamma Ray Background}
\label{gammasec}

To calculate the gamma ray flux from cosmic ray interactions with gas in the outer halo of the Milky Way, we adopt a simple diffusion model:

\begin{equation}
\frac{\partial}{\partial t}\frac{dN_p}{dE_p}(\vec{x},t,E_p)=\vec{\nabla}\cdot [K(E_p) \vec{\nabla} \frac{dN_p}{dE_p}(\vec{x},t,E_p)] + Q(\vec{x},t,E_p), \nonumber
\end{equation}
where $K$ is the diffusion coefficient, and $dN_p/dE_p$ describes the distribution and spectrum of cosmic rays. $Q$ is the source term, which describes the injection rate and spectrum of cosmic rays from the disk.  As the escape time from the disk is small compared to the relevant timescales of the problem, we simply assume that cosmic rays immediately escape the disk and diffuse outward according to the $K$ value chosen for the outer halo. We adopt an energy dependence of the source term given by $Q(E_p) \propto E_p^{-2.4}$ (consistent with the observed cosmic ray spectrum, after accounting for diffusion), and have normalized the source term to produce the observed local density of cosmic rays. We also adopt a time dependence in the source term intended to reflect variations in the star formation rate of the Milky Way:
\begin{equation}
\frac{Q(t)}{Q(0)}= 
\begin{cases}
1 + t/(1\,{\rm Gyr})  & \text{if $t\leq{}2$ Gyr,} \\
3                                 & \text{if 2 Gyr $<t\leq{}$6 Gyr,} \\
3-0.5(t-6\,{\rm Gyr}) & \text{if 6 Gyr $<t\leq{}$10 Gyr.}
\end{cases}
\end{equation}

We consider a diffusion coefficient that is constant throughout the outer halo and with an energy dependance given by $K(E_p) =K_0 E^{0.33}_p$. We find that for a value of $K_0=1.2 \times 10^{29}$ cm$^2$/s, the cosmic ray halo extends out to $\sim$60 kpc, and the cosmic ray density in the region surrounding the disk is about 30\% of the local cosmic ray density. Much smaller values of  $K_0$ are likely inconsistent with measurements of local primary-to-secondary cosmic ray species. We also consider a substantially larger diffusion coefficient of $K_0=4 \times 10^{30}$ cm$^2$/s, which leads to a cosmic ray halo that extends out to several hundred kpc.

From the cosmic ray distribution and the distribution of ionized $\HII$ gas as shown in Fig.~\ref{gas}, we calculate the spectrum of gamma rays produced per volume from pion production:
\begin{equation}
\frac{dN_{\gamma}}{dE_{\gamma}} = 2 \int^{\infty}_{E^{\rm min}_{\pi}(E_{\gamma})} dE_{\pi} \frac{dN_{\pi}}{dE_{\pi}} \frac{1}{\sqrt{E^2_{\pi}-m^2_{\pi}}},
\end{equation}
where $dN_{\pi}/dE_{\pi}$ is the spectrum of neutral pions produced in cosmic ray-gas collisions:
\begin{equation}
\frac{dN_{\pi}}{dE_{\pi}} = 4 \pi\, n_{\rm H} \int^{\infty}_{E^{\rm min}_{p}(E_{\pi})} dE_p J_p(E_p) \frac{d\sigma_{\pi}}{dE_{\pi}}(E_{\pi},E_p).
\end{equation}
Here, $n_{\rm H}$ is the number density of gas (as shown in Fig.~\ref{gas}) and $J_p(E_p)$ is the cosmic ray intensity (per energy). For useful parameterizations of the differential cross section for pion production, see~\cite{pion}.

\begin{figure}[t]
\centering
\includegraphics[angle=0.0,width=3.2in]{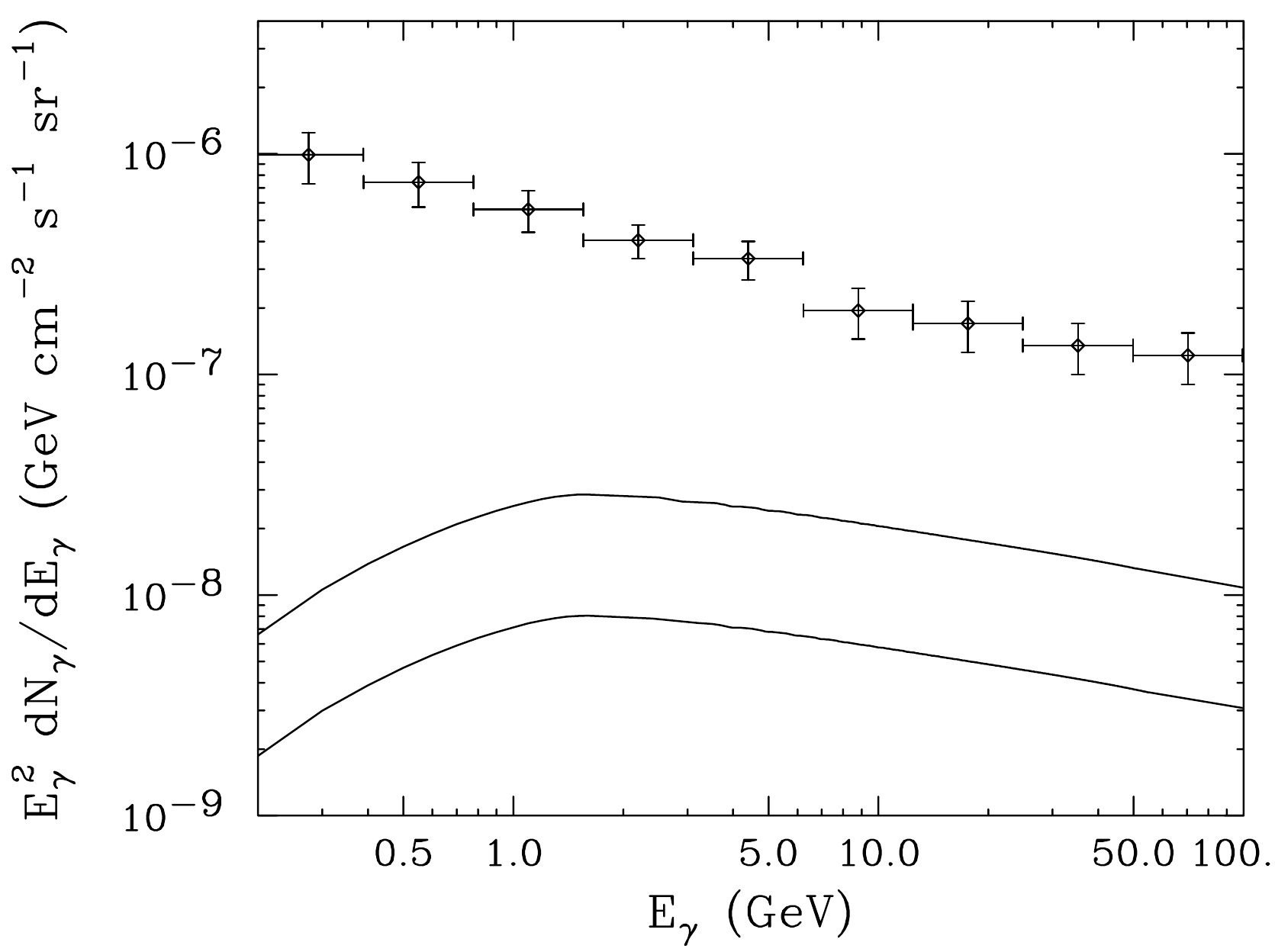}
\caption{The contribution to the high latitude gamma ray background (as measured by the Fermi Gamma-Ray Space Telescope;~\citealt{fermibg}) from cosmic ray interactions with ionized hydrogen ($\HII$) in the outer halo of the Milky Way. Here, we have adopted the gas density shown in Fig.~\ref{gas} and propagated cosmic rays with a diffusion coefficient in the outer halo given by $K(E_p)=1.2\times 10^{29} \,{\rm cm}^2/{\rm s}\, (E_p/{\rm GeV})^{0.33}$ (top) and $K(E_p)=4\times 10^{30} \,{\rm cm}^2/{\rm s}\, (E_p/{\rm GeV})^{0.33}$ (bottom). We expect a contribution at a similar level from gamma rays that are produced by the same mechanism in the $\HII$ halos of all galaxies along a given line-of-sight.}
\label{gamma}
\end{figure}

In Fig.~\ref{gamma}, we show the contribution to the gamma ray background at high Galactic latitudes from cosmic ray interactions with the extended halo of ionized gas. Results are shown for two choices of the diffusion coefficient, which determine how far the cosmic rays have propagated after escaping the disk. For this range of diffusion coefficients, we find that between 3 and 10\% of the isotropic emission observed by Fermi originates from these interactions. 

The spatial extent of the cosmic ray halo represents the most significant uncertainty in our calculation.  The two curves shown in Fig.~\ref{gamma} represent current cosmic ray distributions which fall to half of their density (not including the density within the disk itself) by 60 kpc and by 360 kpc, respectively. For these two cases, half of the observed gamma rays originate from within 18 and 38 kpc, respectively, leading to an approximately isotropic angular distribution of gamma rays, not dissimilar to that predicted in the inverse Compton scenario described by \cite{Keshet:2003xc}. We consider these two cases to represent a reasonable range of possibilities, although gamma ray fluxes higher or lower by a factor of a few are not implausible. 

The overall luminosity of gamma rays originating from cosmic ray interactions with circum-galactic gas is potentially substantial. The large number of cosmic rays in the Galactic halo (see Sec.~\ref{crs}) compensates for the low density of the circum-galactic gas. We can obtain an order of magnitude estimate by integrating the cosmic ray injection rate over the last 10 Gyr and assuming that most of the cosmic rays remain within the virial radius of the Galaxy, where they interact with the circum-galactic gas. Under these assumptions, we obtain a gamma-ray luminosity of $L_\gamma{}(>100\,{\rm MeV})\sim{}9\times{}10^{38}$ erg s$^{-1} \times (\langle{}n_{\rm H}\rangle{}/10^{-4} {\rm cm}^{-3})$, where $\langle{}n_{\rm H}\rangle{}$ is the typical circum-galactic gas density encountered by cosmic rays. For $\langle{}n_{\rm H}\rangle{}\sim{}10^{-4}$ cm$^{-3}$, the gamma-ray luminosity from the circum-galactic gas is comparable with that induced by cosmic rays interacting with the interstellar medium in the Galactic disk \citep{2010ApJ...722L..58S} or that of M31 \citep{2010A&A...523L...2A}. 

The gamma ray luminosity from such interactions in the extended gaseous halos of other galaxies are also expected to contribute to the observed isotropic gamma ray background. We can use the estimate for the circum-galactic gamma ray luminosity given in the previous paragraph to judge whether this contribution is likely to be significant. We scale the gamma ray luminosity of a given galaxy with its stellar mass, which is an indicator of the total number of cosmic rays injected into the halo of ionized gas. Then, using the redshift-dependent stellar mass to halo mass relation by \cite{2011arXiv1110.1420Y} and the halo mass function by \cite{2011ApJ...740..102K}, we estimate the contribution to the local gamma ray flux by integrating over all redshifts and halo masses. We note that only galaxies below $z\sim{}1$ and with $M_{\rm vir}\sim{}10^{11}-10^{13}$ $M_\odot$ are particularly relevant for this calculation. Galaxies in lower mass halos do not contain a sufficient number of cosmic rays to contribute significantly, and galaxies that reside in very massive halos are too rare. 
Overall, we find that the total contribution from extended gaseous halos of all galaxies should account for a few percent of the isotropic gamma ray background.  This contribution is thus similar to that originating in the ionized halo of our own galaxy.

\section{Discussion and Conclusions}
\label{conclusions}

The Fermi Gamma Ray Space Telescope has revealed the presence of very little anisotropy in the diffuse gamma ray background~\citep{2012PhRvD..85h3007A, 2012arXiv1202.5309C}, leading to the disfavoring of blazars and other rare but luminous gamma ray sources as the dominant contributors to this background. Instead, it seems that more faint and numerous sources, such as starburst galaxies, are more likely to make up the bulk of the observed emission.

In this paper, we have considered an alternative contribution to the observed isotropic gamma ray background. In particular, we have shown that the presence of an extended halo of ionized gas ($\HII$) surrounding the Milky Way, as favored by both simulations and observations, can provide an important target for cosmic rays that are no longer confined to the disk, leading to a significant contribution to the high-latitude, diffuse gamma ray flux. Using a distribution of $\HII$ gas derived using a high resolution hydrodynamical simulation and reasonable estimates for the distribution of cosmic rays in the outer halo of the Milky Way, we estimate that pion producing interactions taking place within the surrounding several tens of kiloparsecs from the Galactic disk may account for on the order of 3-10\% of the isotropic gamma ray background as observed by the Fermi Gamma Ray Space Telescope. In addition, the corresponding gamma ray emission originating in the $\HII$ halos of galaxies along a given line-of-sight should contribute at a similar level.

In addition to the gamma ray background under consideration here, it is interesting to contemplate whether an extended halo of circum-galactic gas may be connected in any way to the significant excess of isotropic radio emission~\citep{arcadeinterpretation} that has been observed by several groups~\citep{haslam, reich, roger, arcade,maeda}. In particular, one could imagine energetic electrons and positrons being generated in cosmic ray collisions with the circum-galactic gas, resulting in the production of an approximately isotropic background of radio synchrotron. With little information pertaining to the magnetic fields present in the region of the outer halo, however, it is difficult to make quantitative estimates of such a signal. 

\smallskip

We would like to thank Anders Pinzke and Surhud More for valuable input and discussions. This work has been supported by the U.S. Department of Energy. Fermilab is operated by Fermi Research Alliance, LLC under contract No. DE-AC02-07CH11359 with the United States Department of Energy. 

\bibliographystyle{apj}

\begin{thebibliography}{}

\bibitem[Abazajian et al.(2011)]{2011PhRvD..84j3007A} Abazajian, K.~N., 
Blanchet, S., \& Harding, J.~P.\ 2011, \prd, 84, 103007 
 
\bibitem[Abdo et al.(2010)]{fermibg} Abdo, A.~A., Ackermann, 
M., Ajello, M., et al.\ 2010, Physical Review Letters, 104, 101101
  
\bibitem[Abdo et al.(2010)]{2010A&A...523L...2A} Abdo, A.~A., Ackermann, M., Ajello, M., et al.\ 2010, \aap, 523, L2 
  
\bibitem[Ackermann et al.(2012)]{2012PhRvD..85h3007A} Ackermann, M., 
Ajello, M., Albert, A., et al.\ 2012, \prd, 85, 083007 

\bibitem[Anderson \& Bregman(2010)]{2010ApJ...714..320A} Anderson, M.~E., \& Bregman, J.~N.\ 2010, \apj, 714, 320 

\bibitem[Anderson \& Bregman(2011)]{2011ApJ...737...22A} Anderson, M.~E., \& Bregman, J.~N.\ 2011, \apj, 737, 22 

\bibitem[Belikov 
\& Hooper(2010)]{2010PhRvD..81d3505B} Belikov, A.~V., \& Hooper, D.\ 2010, \prd, 81, 043505 

\bibitem[{{Bertschinger}(2001)}]{2001ApJS..137....1B}
{Bertschinger}, E. 2001, \apjs, 137, 1

\bibitem[Besla et al.(2012)]{2012MNRAS.421.2109B} Besla, G., Kallivayalil, 
N., Hernquist, L., et al.\ 2012, \mnras, 421, 2109 

\bibitem[Bland-Hawthorn et al.(2007)]{2007ApJ...670L.109B} Bland-Hawthorn, 
J., Sutherland, R., Agertz, O., \& Moore, B.\ 2007, \apj, 670, L109 

\bibitem[Blattnig et al.(2000)]{pion} Blattnig, S.~R., 
Swaminathan, S.~R., Kruger, A.~T., Ngom, M., 
\& Norbury, J.~W.\ 2000, \prd, 62, 094030 

\bibitem[Blitz \& Robishaw(2000)]{2000ApJ...541..675B} Blitz, L., \& Robishaw, T.\ 2000, \apj, 541, 675 

\bibitem[Bregman(2007)]{2007ARA&A..45..221B} Bregman, J.~N.\ 2007, Annu. Rev. Astron. Astrophys., 45, 221 

\bibitem[Br{\"u}ns et 
al.(2000)]{2000A&A...357..120B} Br{\"u}ns, C., Kerp, J., Kalberla, P.~M.~W., \& Mebold, U.\ 2000, \aap, 357, 120 

\bibitem[Casanova et al.(2007)]{2007ApJ...656..306C} Casanova, S., Dingus, 
B.~L., \& Zhang, B.\ 2007, \apj, 656, 306 

\bibitem[Combes et 
al.(2009)]{2009A&A...503...73C} Combes, F., Baker, A.~J., Schinnerer, E., et al.\ 2009, \aap, 503, 73 

\bibitem[Crain et al.(2010)]{2010MNRAS.407.1403C} Crain, R.~A., McCarthy, I.~G., Frenk, C.~S., Theuns, T., \& Schaye, J.\ 2010, \mnras, 407, 1403 

\bibitem[Cuoco et al.(2012)]{2012arXiv1202.5309C} Cuoco, A., Komatsu, E., 
\& Siegal-Gaskins, J.\ 2012, arXiv:1202.5309 

\bibitem[{{Feldmann} {et~al.}(2011){Feldmann}, {Gnedin}, \&
 {Kravtsov}}]{2011ApJ...732..115F}
{Feldmann}, R., {Gnedin}, N.~Y., \& {Kravtsov}, A.~V. 2011, \apj, 732, 115

\bibitem[Feldmann et al.(2012)]{2012arXiv1204.3910F} Feldmann, R., Gnedin, 
N.~Y., \& Kravtsov, A.~V.\ 2012, arXiv:1204.3910 

\bibitem[Fichtel et al.(1978)]{sas} Fichtel, C.~E., 
Simpson, G.~A., \& Thompson, D.~J.\ 1978, \apj, 222, 833 

\bibitem[Fixsen et al.(2011)]{arcade} Fixsen, D.~J., Kogut, 
A., Levin, S., et al.\ 2011, \apj, 734, 5 

\bibitem[Forman et al.(1985)]{1985ApJ...293..102F} Forman, W., Jones, C., 
\& Tucker, W.\ 1985, \apj, 293, 102 

\bibitem[{{Gnedin} \& {Abel}(2001)}]{2001NewA....6..437G}
{Gnedin}, N.~Y., \& {Abel}, T. 2001, New Astronomy, 6, 437

\bibitem[{{Gnedin} {et~al.}(2009){Gnedin}, {Tassis}, \&
 {Kravtsov}}]{2009ApJ...697...55G}
{Gnedin}, N.~Y., {Tassis}, K., \& {Kravtsov}, A.~V. 2009, \apj, 697, 55

\bibitem[{{Gnedin} \& {Kravtsov}(2011)}]{2011ApJ...728...88G}
{Gnedin}, N.~Y., \& {Kravtsov}, A.~V. 2011, \apj, 728, 88

\bibitem[Gnedin(2011)]{2011arXiv1108.2271G} Gnedin, N.~Y.\ 2011, arXiv:1108.2271 

\bibitem[Grcevich \& Putman(2009)]{2009ApJ...696..385G} Grcevich, J., \& Putman, M.~E.\ 2009, \apj, 696, 385 

\bibitem[Guzm{\'a}n et 
al.(2011)]{maeda} Guzm{\'a}n, A.~E., May, J., Alvarez, H., \& Maeda, K.\ 2011, \aap, 525, A138 

\bibitem[Haan et al.(2008)]{2008AJ....135..232H} Haan, S., Schinnerer, E., Mundell, C.~G., Garc{\'{\i}}a-Burillo, S., \& Combes, F.\ 2008, \aj, 135, 232 

\bibitem[Harding 
\& Abazajian(2012)]{Harding:2012sa} Harding, J.~P., \& Abazajian, K.~N.\ 2012, arXiv:1204.3870 

\bibitem[Haslam et 
al.(1981)]{haslam} Haslam, C.~G.~T., Klein, U., Salter, C.~J., et al.\ 1981, \aap, 100, 209 

\bibitem[Hickox 
\& Markevitch(2006)]{2006ApJ...645...95H} Hickox, R.~C., \& Markevitch, M.\ 2006, \apj, 645, 95 

\bibitem[Humphrey et al.(2011)]{2011ApJ...729...53H} Humphrey, P.~J., 
Buote, D.~A., Canizares, C.~R., Fabian, A.~C., 
\& Miller, J.~M.\ 2011, \apj, 729, 53 

\bibitem[Humphrey et al.(2012)]{2012arXiv1204.3095H} Humphrey, P.~J., 
Buote, D.~A., O'Sullivan, E., \& Ponman, T.~J.\ 2012, arXiv:1204.3095 


\bibitem[Kalashev et al.(2009)]{Kalashev:2007sn} Kalashev, O.~E., 
Semikoz, D.~V., \& Sigl, G.\ 2009, \prd, 79, 063005 

\bibitem[{{Katz}(1991)}]{1991ApJ...368..325K}
{Katz}, N. 1991, \apj, 368, 325

\bibitem[Keshet et al.(2003)]{2003ApJ...585..128K} Keshet, U., Waxman, E., 
Loeb, A., Springel, V., \& Hernquist, L.\ 2003, \apj, 585, 128 

\bibitem[Keshet et al.(2004)]{Keshet:2003xc} Keshet, U., Waxman, E., 
\& Loeb, A.\ 2004, \jcap, 4, 6 

\bibitem[Klypin et al.(2011)]{2011ApJ...740..102K} Klypin, A.~A., Trujillo-Gomez, S., \& Primack, J.\ 2011, \apj, 740, 102 

\bibitem[Kneiske 
\& Mannheim(2005)]{2005AIPC..745..578K} Kneiske, T.~M., \& Mannheim, K.\ 2005, High Energy Gamma-Ray Astronomy, 745, 578 

\bibitem[{{Kravtsov} {et~al.}(1997){Kravtsov}, {Klypin}, \&
 {Khokhlov}}]{1997ApJS..111...73K}
{Kravtsov}, A.~V., {Klypin}, A.~A., \& {Khokhlov}, A.~M. 1997, \apjs, 111, 73

\bibitem[{{Kravtsov} {et~al.}(2002){Kravtsov}, {Klypin}, \&
 {Hoffman}}]{2002ApJ...571..563K}
{Kravtsov}, A.~V., {Klypin}, A., \& {Hoffman}, Y. 2002, \apj, 571, 563

\bibitem[Kuntz 
\& Snowden(2000)]{2000ApJ...543..195K} Kuntz, K.~D., \& Snowden, S.~L.\ 2000, \apj, 543, 195 

\bibitem[Loeb 
\& Waxman(2000)]{2000Natur.405..156L} Loeb, A., \& Waxman, E.\ 2000, \nat, 405, 156 

\bibitem[Lumb et 
al.(2002)]{2002A&A...389...93L} Lumb, D.~H., Warwick, R.~S., Page, M., \& De Luca, A.\ 2002, \aap, 389, 93 

\bibitem[Maller \& Bullock(2004)]{2004MNRAS.355..694M} Maller, A.~H., \& Bullock, J.~S.\ 2004, \mnras, 355, 694 

\bibitem[Nicastro et al.(2003)]{2003Natur.421..719N} Nicastro, F., Zezas, 
A., Elvis, M., et al.\ 2003, \nat, 421, 719 

\bibitem[Peek et al.(2007)]{2007ApJ...656..907P} Peek, J.~E.~G., Putman, 
M.~E., McKee, C.~F., Heiles, C., \& Stanimirovi{\'c}, S.\ 2007, \apj, 656, 907 

\bibitem[Putman et al.(2003)]{2003ApJ...586..170P} Putman, M.~E., 
Staveley-Smith, L., Freeman, K.~C., Gibson, B.~K., 
\& Barnes, D.~G.\ 2003, \apj, 586, 170 

\bibitem[Putman et al.(2011)]{2011MNRAS.418.1575P} Putman, M.~E., Saul, 
D.~R., \& Mets, E.\ 2011, \mnras, 418, 1575 

\bibitem[Quilis \& Moore(2001)]{2001ApJ...555L..95Q} Quilis, V., \& Moore, B.\ 2001, \apjl, 555, L95 

\bibitem[Rasmussen et al.(2003)]{2003ASSL..281..109R} Rasmussen, A., Kahn, 
S.~M., \& Paerels, F.\ 2003, ASSL Conference Proceedings Vol. 281, 109
The IGM/Galaxy Connection.~The Distribution of Baryons at z=0, 

\bibitem[Rasmussen et al.(2009)]{2009ApJ...697...79R} Rasmussen, J., Sommer-Larsen, J., Pedersen, K., et al.\ 2009, \apj, 697, 79 

\bibitem[Reich 
\& Reich(1986)]{reich} Reich, P., \& Reich, W.\ 1986, \aaps, 63, 205 

\bibitem[Roger et 
al.(1999)]{roger} Roger, R.~S., Costain, C.~H., Landecker, T.~L., \& Swerdlyk, C.~M.\ 1999, \aaps, 137, 7 

\bibitem[Seiffert et al.(2011)]{arcadeinterpretation} Seiffert, M., Fixsen, 
D.~J., Kogut, A., et al.\ 2011, \apj, 734, 6 

\bibitem[Sembach et al.(2003)]{2003ApJS..146..165S} Sembach, K.~R., Wakker, 
B.~P., Savage, B.~D., et al.\ 2003, \apjs, 146, 165 

\bibitem[Sommer-Larsen(2006)]{2006ApJ...644L...1S} Sommer-Larsen, J.\ 2006, \apjl, 644, L1 

\bibitem[{{Spergel} {et~al.}(2003){Spergel}, {Verde}, {Peiris}, {Komatsu},
 {Nolta}, {Bennett}, {Halpern}, {Hinshaw}, {Jarosik}, {Kogut}, {Limon},
 {Meyer}, {Page}, {Tucker}, {Weiland}, {Wollack}, \&
 {Wright}}]{2003ApJS..148..175S}
{Spergel}, D.~N., {Verde}, L., {Peiris}, H.~V., {Komatsu}, E., {Nolta}, M.~R.,
 {Bennett}, C.~L., {Halpern}, M., {Hinshaw}, G., {Jarosik}, N., {Kogut}, A.,
 {Limon}, M., {Meyer}, S.~S., {Page}, L., {Tucker}, G.~S., {Weiland}, J.~L.,
 {Wollack}, E., \& {Wright}, E.~L. 2003, \apjs, 148, 175

\bibitem[Sreekumar et al.(1998)]{egret} Sreekumar, P., 
Bertsch, D.~L., Dingus, B.~L., et al.\ 1998, \apj, 494, 523 

\bibitem[Stanimirovi{\'c} et al.(2002)]{2002ApJ...576..773S} 
Stanimirovi{\'c}, S., Dickey, J.~M., Kr{\v c}o, M., 
\& Brooks, A.~M.\ 2002, \apj, 576, 773 

\bibitem[Stecker 
\& Salamon(1996)]{1996ApJ...464..600S} Stecker, F.~W., \& Salamon, M.~H.\ 1996, \apj, 464, 600 

\bibitem[Strong et al.(2007)]{Strong:2007nh} Strong, A.~W., 
Moskalenko, I.~V., 
\& Ptuskin, V.~S.\ 2007, Annual Review of Nuclear and Particle Science, 57, 285 

\bibitem[Strong et al.(2010)]{2010ApJ...722L..58S} Strong, A.~W., Porter, T.~A., Digel, S.~W., et al.\ 2010, \apjl, 722, L58 

\bibitem[Thompson et al.(2007)]{Thompson:2006qd} Thompson, T.~A., 
Quataert, E., \& Waxman, E.\ 2007, \apj, 654, 219 

\bibitem[Toft et al.(2002)]{2002MNRAS.335..799T} Toft, S., Rasmussen, J., Sommer-Larsen, J., \& Pedersen, K.\ 2002, \mnras, 335, 799 

\bibitem[Ullio et al.(2002)]{2002PhRvD..66l3502U} Ullio, P., Bergstr{\"o}m, 
L., Edsj{\"o}, J., \& Lacey, C.\ 2002, \prd, 66, 123502 

\bibitem[Weiner \& Williams(1996)]{1996AJ....111.1156W} Weiner, B.~J., \& Williams, T.~B.\ 1996, \aj , 111, 1156 

\bibitem[Yang et al.(2011)]{2011arXiv1110.1420Y} Yang, X., Mo, H.~J., van den Bosch, F.~C., Zhang, Y., \& Han, J.\ 2011, arXiv:1110.1420 

\end{thebibliography}

\end{document}